\documentclass[journal,comsoc]{IEEEtran}

\usepackage{amsmath}
\usepackage{bm}
\usepackage{xcolor}
\usepackage{graphicx}
\usepackage{caption,float}
\usepackage{subcaption}
\usepackage[american]{circuitikz}
\usepackage{mathtools,amssymb,lipsum}

\usepackage{cuted}
\begin{document}
\title{A Generalized Theory of Power}

\author{Louis L. Scharf, \IEEEmembership{Life Fellow, IEEE} and Dongliang Duan, \IEEEmembership{Senior Member, IEEE}
\thanks{Louis L. Scharf is with the Department of Mathematics and the Department of Statistics, Colorado State University, Fort Collins, CO 80521. Email: Louis.Scharf{\rm\char64}colostate.edu.}

\thanks{Dongliang Duan is with the Department of Electrical and Computer Engineering, University of Wyoming, Laramie, WY 82071. Email: dduan{\rm\char64}uwyo.edu.}

}
\maketitle

\begin{abstract}
The complex representation of real-valued instantaneous power may be written as the sum of two complex powers, one Hermitian and the other non-Hermitian, or complementary. A virtue of this representation is that it consists of a power triangle rotating around a fixed phasor, thus clarifying what should be meant by the power triangle.  The in-phase and quadrature components of complementary power encode for active and non-active power. When instantaneous power is defined for a Thevenin equivalent circuit, these  are time-varying real and reactive power components.   These claims hold for sinusoidal voltage and current, and for non-sinusoidal voltage and current.  Spectral representations of Hermitian, complementary, and instantaneous power show that, frequency-by-frequency, these powers behave exactly as they behave in the single frequency sinusoidal case. Simple hardware diagrams show how instantaneous active  and non-active power may be extracted from metered voltage and current, even in certain non-sinusoidal cases.  
\end{abstract}

\section{Introduction}
Definitions and interpretations of active  and non-active power in non-sinusoidal systems have engaged the interest of many important and influential investigators \cite{Bud1927}-\cite{Mun}.  Nonetheless, there remain issues to be resolved. 

The first issue is the resolution of sinusoidal instantaneous power into its constituents, the second is the extension of this resolution to non-sinusoidal systems, and  the third is the 
display of these constituents to the power system engineer who wishes to track power system performance in real time.

In this paper we begin with the standard resolution of real-valued instantaneous power into {\em average, active, and non-active  power},\footnote{Throughout, we shall follow the terminology of IEEE Standard 1459, 2010.} and offer an easily interpreted complex representation of this real-valued  power. In a complex representation of real-valued instantaneous power,  a complex {\em power triangle}, rotates around a fixed complex phasor. In a Thevenin equivalent circuit, the orthogonal components of this power triangle  are in-phase and quadrature components that may be identified with power exchange between reactive components and resistive components. 
In any particular circuit this  power accounts for the rate at which reactive energy is exchanged between electric fields that sustain voltage, magnetic fields that sustain current, and heat, light, or mass movement that is produced in resistive  components. This interpretation provides an alternative view of the  power triangle. 

We conclude this beginning section with a definition of {\em positive power} and {\em negative power}. These resolve instantaneous power into positive power delivered from the energy source to the energy destination, and the negative power  returned to the source from the destination. During each period of sinusoidal oscillation, the real-valued instantaneous power is negative over short periods of time, during which the rate of exchange of energy from electric and magnetic fields to real elements exceeds the rate at which heat can be dissipated, lights illuminated, or shafts turned. This excess is returned to the source.  The averages of each of the positive and negative power components show the role that the {\em power angle} plays in determining negative power.

Then arises the question, ``how much of this analysis can be extended to non-sinusoidal systems?" In other words, can  complex phasors, which are special cases of {\em complex analytic signals}, be extended to more general complex analytic signals? And if so, can real-valued instantaneous  power still be represented by complex power, in which a slowly re-shaped complex  power triangle rotates around a slowly varying complex analytic signal? The answer is `yes' in any non-sinusoidal system for which Bedrosian's theorem applies. In the context of power systems, these are systems in which the non-sinusoidal signal consists of slowly-varying amplitude and phase modulation of a sinusoidal carrier. 

Throughout the paper, the line of argument is this: begin with real-valued instantaneous power, give it a complex representation, and use this complex representation to develop intuition, and derive equations and  diagrams for extracting the components of real-valued  power. It is to be emphasized that power, as we define it, is always real-valued. Nonetheless, complex representations of this real-valued  power bring intuition and economies of  reasoning. In particular, complex power produces a geometric picture that is not evident in the resolution of this complex power onto the real axis of the Argand plane, where one finds real-valued  power. Moreover, it is found that this geometric picture is most illuminating when complex power is resolved into a sum of {\em Hermitian complex power} and {\em non-Hermitian complementary complex power}. These points will be clarified in due course.
 
The rest of the paper is organized as follows. In Section II, we will discuss the instantaneous, average, in-phase and quadrature power in sinusoidal systems. Specifically, we will show the interpretation of all the power quantities in a Thevenin equivalent circuit. We will introduce the Hilbert transform and Bedrosian's Theorem in Section III, which leads to our complex representation of instantaneous power for non-sinusoidal signals discussed in Section IV. Interpretations of our generalized power theory in a Thevenin equivalent circuit with time-varying impedances are presented in Section V. Then, a spectral theory of average power is introduced in Section VI with hardware diagrams for extracting components of instantaneous power given in Section VII. Concluding remarks are given in Section VIII.

\section{Instantaneous, Average, In-Phase, and Quadrature Power in Sinusoidal Systems}

Let us begin with a simple example to motivate our investigations. Define the real voltage $v(t)=V\cos(\omega_0t+\theta)$ and the real  current $i(t)=I\cos(\omega_0t+\phi)$. The corresponding RMS voltage and current are $V/\sqrt{2}$ and $I/\sqrt{2}$. The {\em apparent power} is defined to be the product $VI/2$. 

The real-valued {\em instantaneous power} is defined to be $p(t)=v(t)i(t)$, which may be written 
\begin{align}
\label{eq:instpower1}
p_{vi}(t)&=v(t)i(t)\nonumber\\&=\frac{VI}{2}\cos(\theta-\phi)+\frac{VI}{2}\cos(2\omega_0t+\theta+\phi),\\
\label{eq:instpower2}
&=\frac{VI}{2}\cos(\theta-\phi)+\frac{VI}{2}\cos(\theta-\phi)\cos(2\omega_0 t+2\phi) \nonumber\\
&~~~~~-\frac{VI}{2}\sin(\theta-\phi)\sin(2\omega_0 t+2\phi).
\end{align}
Equation \eqref{eq:instpower1} shows real-valued instantaneous power to oscillate at frequency $2\omega_0$ and phase delay $\frac{-(\theta+\phi)}{2\omega_0}$ around the {\em average power} $P_{vi}=\frac{VI}{2}\cos(\theta-\phi)$. 
This is a scaling of {\em apparent power} $\frac{VI}{2}$ by the {\em power factor} $\cos(\theta-\phi)$. The sum of the first two terms in Eq. (\ref{eq:instpower2})  is {\em active power}. It is always non-negative. The third term is time-varying {\em non-active power}. In a Thevenin equivalent circuit, active power is {\em real power} and non-active power is {\em reactive power}.  

\subsection{Complex Representation of Instantaneous Power}
There is a complex representation of real-valued instantaneous power, based on a phasor, or complex analytic, representation of real signals:
{\begin{align}
&p_{vi}(t)=\makebox{Re}\left\{\frac{VI}{2}e^{j(\theta-\phi)}+\frac{VI}{2}e^{j(\theta+\phi)}e^{j2\omega_0t}\right\} \label{eq:complexinstpower1}\\
=&\makebox{Re}\left\{\frac{VI}{2}e^{j(\theta-\phi)}+\left[\frac{VI}{2}\cos(\theta-\phi)+ j\frac{VI}{2}\sin(\theta-\phi)\right]e^{j(2\omega_0 t+2\phi)}\right\}\label{eq:complexinstpower2}\\
=&\frac{VI}{2}\cos(\theta-\phi)+\frac{VI}{2}\cos(\theta-\phi)\cos(2\omega_0 t+2\phi)
\nonumber\\&-\frac{VI}{2}\sin(\theta-\phi)\sin(2\omega_0t+2\phi).\label{eq:complexinstpower3}
\end{align}}

\begin{figure}[h!]
  \centering
  \begin{subfigure}[b]{0.4\textwidth}
         \centering
         \includegraphics[width=\textwidth]{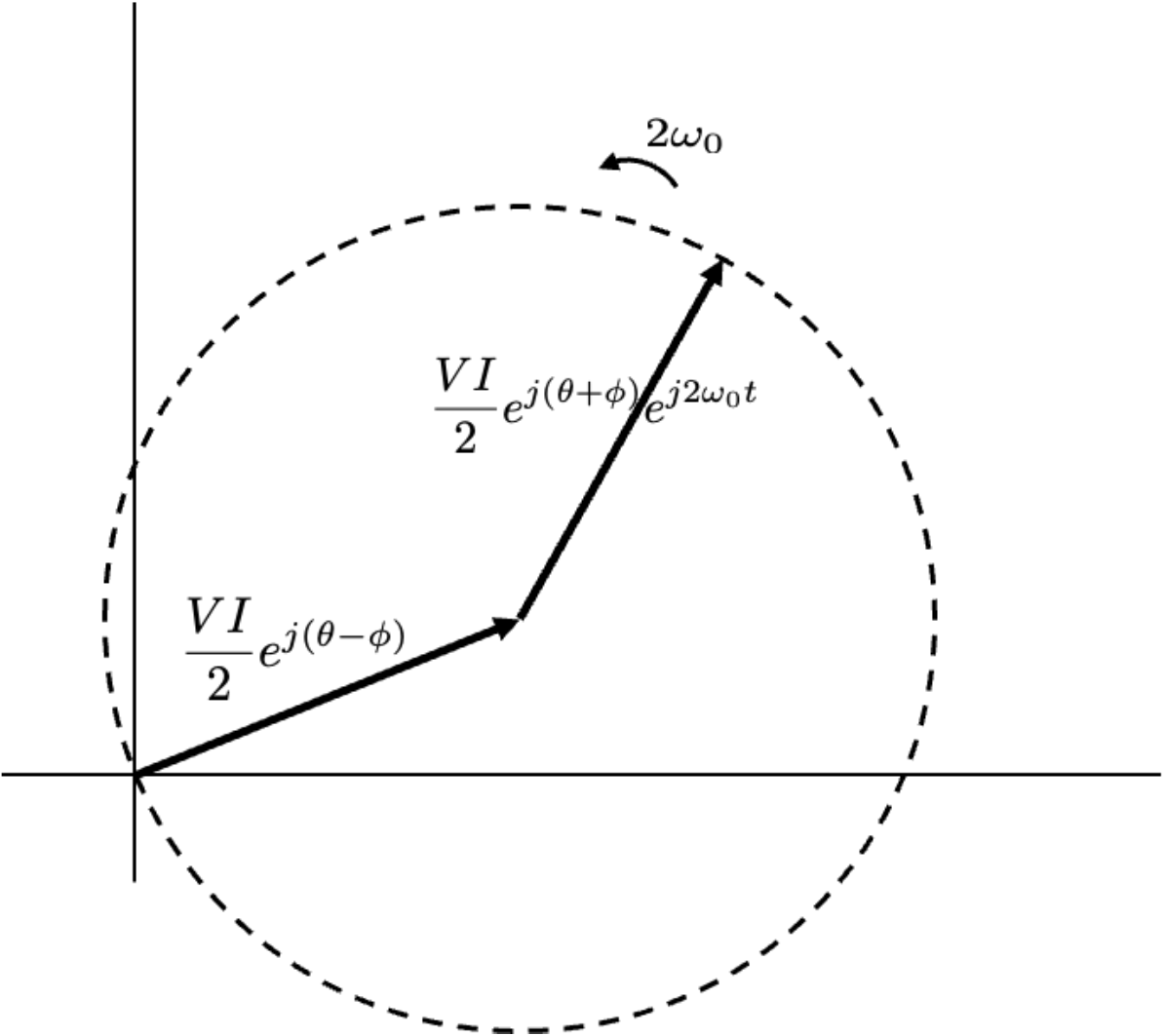}
         \caption{}
     \end{subfigure}
     \hfill
     \begin{subfigure}[b]{0.5\textwidth}
         \centering
         \includegraphics[width=\textwidth]{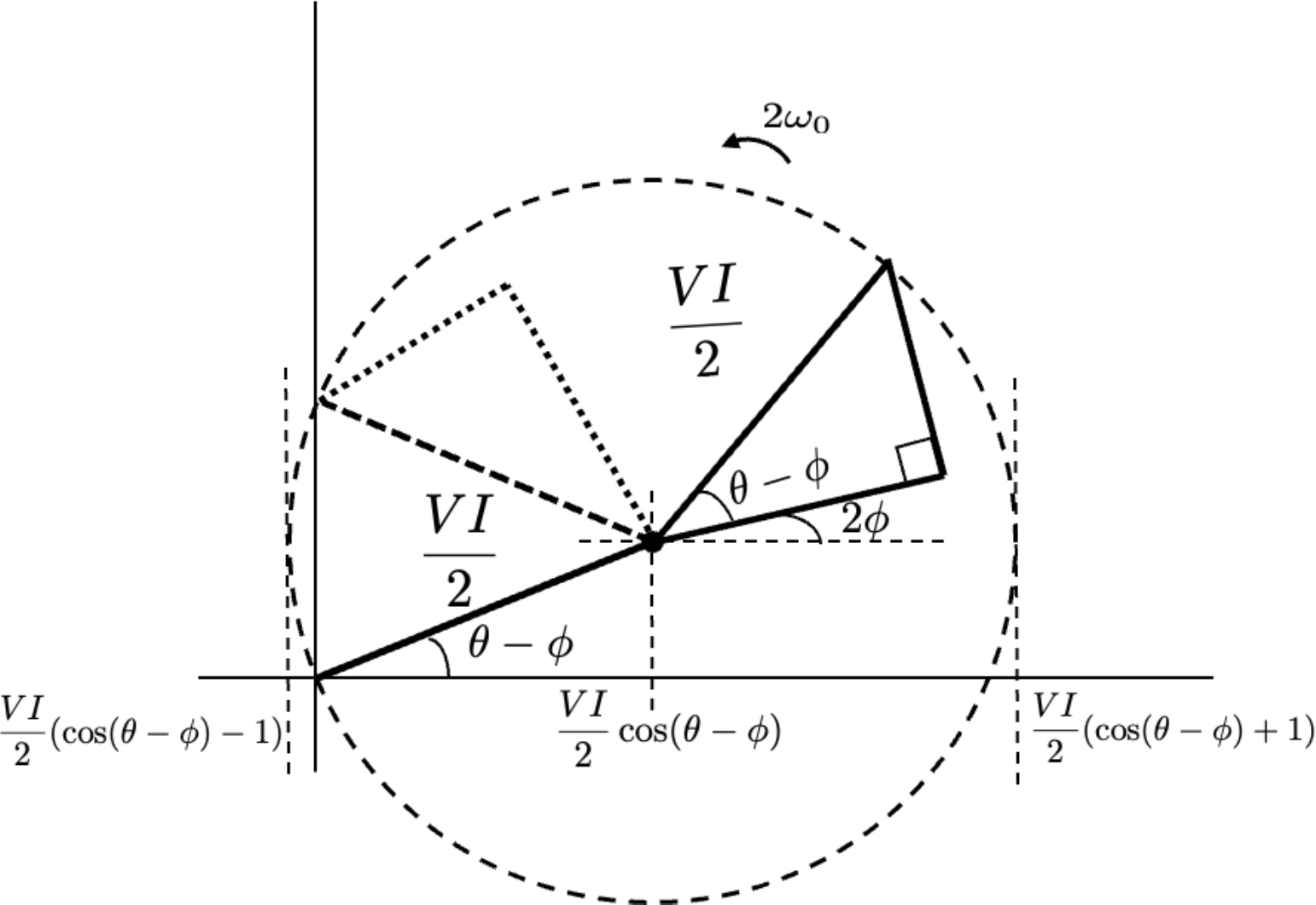}
         \caption{}
        \label{fig:1b}
     \end{subfigure}
    \caption{Phasor illustrations of the instantaneous power.}\label{fig:apparent}
\end{figure}

The term within the $\makebox{Re}$ operator of Eq. \eqref{eq:complexinstpower1} may be termed {\em complex instantaneous power}. It consists of two terms: $(VI/2)e^{j(\theta-\phi)}$ and $(VI/2)e^{j(\theta+\phi)}e^{j2\omega_0t}$. The first is a stationary phasor or analytic signal representation of time-invariant {\em average power} and the second is a rotating phasor or analytic representation of the  time varying component of instantaneous power. In Eq. \eqref{eq:complexinstpower2}, the rotating phasor is re-written as a rotation of the stationary phasor $\frac{VI}{2}e^{j(\theta-\phi)}=\frac{VI}{2}\cos(\theta-\phi)+ j\frac{VI}{2}\sin(\theta-\phi)$. In this Cartesian representation, the components $\left(\frac{VI}{2}\cos(\theta-\phi), \frac{VI}{2}\sin(\theta-\phi)\right)$ are in-phase and quadrature components of the phasor, or adjacent and opposite sides of a right triangle. In this right triangle, the Pythagoren Theorem shows that the hypotenuse has length $\frac{VI}{2}$, which is commonly called {\em apparent power}.  Figure \ref{fig:apparent}(a) is the  complex representation of  instantaneous power given in Eq. (\ref{eq:complexinstpower1}). It consists of a  complex phasor, rotating around a fixed phasor. Instantaneous power is read off the diagram as the real part of the complex instantaneous power. Figure 1(b) resolves the rotating phasor into its in-phase and quadrature components, as in Eq. (\ref{eq:complexinstpower2}). This figure is illuminating because it shows the complex representation of real-valued instantaneous power to consist of a {\em power triangle} rotating around the tip of a fixed phasor. The power triangle is a right triangle, defined by in-phase and quadrature components. In this right triangle, the square of apparent power is the sum of squares of active and non-active powers: $(VI/2)^2=(VI/2)^2\cos^2(\theta-\phi)+ (VI/2)^2\sin^2(\theta-\phi)$.  In IEEE Standard 1459, this Pythagorean identity is written, $S^2=P^2+Q^2$, a notation which is justified by the complex representation 
$\frac{VI}{2}e^{j(\theta-\phi)}=\frac{VI}{2}\cos(\theta-\phi)+ j\frac{VI}{2}\sin(\theta-\phi)$. In a Thevenin equivalent circuit, the square of apparent power is the sum of squares of real and reactive powers. In summary, the power triangle is triangle of Fig. \ref{fig:1b} that rotates around a fixed phasor.

\subsection{Average Power, Positive Power, and Negative Power}
Real-valued instantaneous  power is bounded below by $\frac{VI}{2}[\cos(\theta-\phi)-1]$ and above by $\frac{VI}{2}[\cos(\theta-\phi)+1]$. So it can be negative! This motivates the definition of {\em positive power} $p^+(t)$ and {\em negative power} $p^-(t)$:
\begin{align}
\label{eq:posplusnegpower}
p_{vi}(t)=&p_{vi}^{+}(t)+p_{vi}^{-}(t),\\
\label{eq:posandnegpower}
p_{vi}^{+}(t)=\makebox{max}(p_{vi}(t), 0) &\makebox{ and } p_{vi}^{-}(t)=\makebox{min}(p_{vi}(t), 0)
\end{align}
For $T=\frac{2\pi}{\omega_0}$, the average of negative power over a period is
\begin{equation}
P_{vi}^-=\frac{1}{T}\int_{0}^T p^-(t)dt=-\frac{VI}{2}\left[\frac{\sin({\theta-\phi})}{\pi}-\frac{\theta-\phi}{\pi}\cos(\theta-\phi)\right]\le 0\;,
\end{equation}
and the average of positive power is
\begin{equation}
P_{vi}^+=\frac{VI}{2}\cos(\theta-\phi)+\frac{VI}{2}\left[\frac{\sin({\theta-\phi})}{\pi}-\frac{\theta-\phi}{\pi}\cos(\theta-\phi)\right]\ge 0\;.
\end{equation}
These results show that the {\em power angle} $\theta-\phi$ determines not only the fraction of apparent power 
that is delivered to the load, but it shows that this delivered power is   the sum of average positive power and average negative power:
\begin{align}
P_{vi}=\frac{1}{T}\int_0^T\,p_{vi}(t)dt=P_{vi}^{+} + P_{vi}^{-}=\frac{VI}{2} \cos(\theta-\phi)
\end{align} 
This result clarifies that the difference between apparent power and delivered power is accounted for by the power returned from the load to the source. 

The period of oscillation of instantaneous power is $T=2\pi/2\omega_0$, and the fraction of this period during which power is delivered is $1-(\theta-\phi)/\pi$. So even though average power $\frac{VI}{2}\cos(\theta-\phi)$ decreases trigonometrically  with the power angle $\theta-\phi$, the fraction of time that power is delivered decreases linearly with the power angle. The plot of Figure 2 illustrates typical voltage and currents waveforms for which real-valued instantaneous power is negative over a fraction of a period. The resolution of instantaneous power as a sum of {\em plus power} and {\em negative power} 
requires a  non-linear decomposition of power. 



\begin{figure}[h!]
  \centering
  \includegraphics[width=\linewidth]{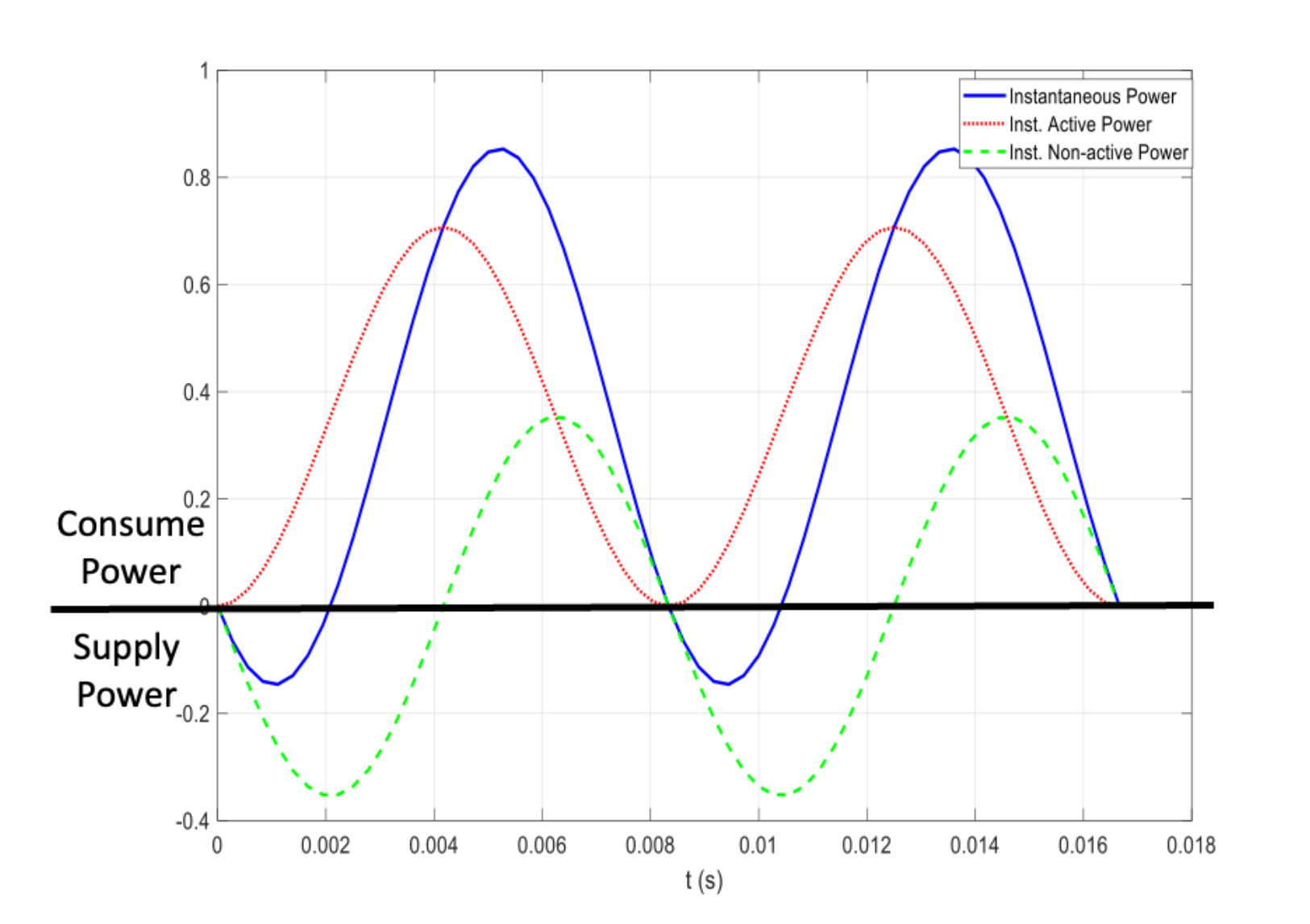}
  \caption{\small Instantaneous power, instantaneous active power, and instantaneous non-active power.}
  \label{fig:inst}
\end{figure}



\begin{figure}[h!]
  \begin{center}
    \begin{circuitikz}
      \draw (0,0)
      to[open] (0,4) 
      to[short, i=$I$] (4,4)
      to[generic=$Z$] (4,0) 
      to[short] (0,0);
      \draw (0,4)
      to[open, v=$V$] (0,0);
    \end{circuitikz}
    \caption{A Thevenin equivalent circuit.}\label{fig:Thev}
  \end{center}
\end{figure}
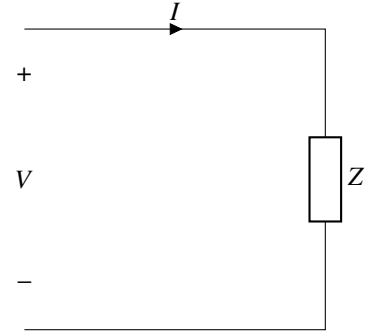

\subsection{Interpretations in a Thevenin Equivalent Circuit} 
When can active power be termed real power, and non-active power be termed reactive power? 

Consider the Thevenin equivalent  circuit in Figure \ref{fig:Thev}, consisting of a complex impedance $Z$ in series with a sinusoidal voltage source $v(t)=\makebox{Re}\{Ve^{j\theta}e^{j\omega_0 t} \}=V\cos(\omega_0 t+\theta)$. 
The sinusoidal current is 
$i(t)=\makebox{Re}\{Ie^{j\phi}e^{j\omega_0 t} \}=I\cos(\omega_0 t+\phi)$. The circuit constraint is 
$Ve^{j\theta}=ZIe^{j\phi}$, where 
the complex impedance is $Z=\frac{V}{I}e^{j(\theta-\phi)}$. The Cartesian representation of this complex impedance is $Z=R+jX$,  where $R=\frac{V}{I}\cos(\theta-\phi)$ is {\em real impedance} and $X=\frac{V}{I}\sin(\theta-\phi)$ is {\em reactive impedance}.  
The term $\frac{VI}{2}e^{j(\theta-\phi)}$ may be written $\frac{ZI^2}{2}$, the term $\frac{VI}{2}\cos(\theta-\phi)$ may  be written $\frac{RI^2}{2}$ and the term $\frac{VI}{2}\sin(\theta-\phi)$ may be 
written $\frac{XI^2}{2}$. With these identities, the complex representation of real-valued instantaneous power in Eqs. \eqref{eq:complexinstpower1} and  \eqref{eq:complexinstpower2} may be written
\begin{align}
\label{eq:inst1}
p_{vi}(t)=&\makebox{Re}\left\{\frac{ZI^2}{2}+\frac{ZI^2}{2}e^{j2\omega_0t}\right\} \\
\label{eq:inst2}
=&\makebox{Re}\left\{\frac{ZI^2}{2}+ \big [\frac{RI^2}{2}+ j\frac{XI^2}{2}\big  ]e^{j(2\omega_0 t+2\phi)}\right\},\\
\label{eq:inst3}
=&\frac{RI^2}{2}\cos(\theta-\phi)+\frac{RI^2}{2}\cos(2\omega_0 t+2\phi)-\frac{XI^2}{2}\sin(2\omega_0 t+2\phi).
\end{align}
Eq. \eqref{eq:inst1} shows that the magnitude of each phasor in the  complex representation of real-valued instantaneous power is determined by $\frac{ZI^2}{2}$. Eq. \eqref{eq:inst2} shows the in-phase component of the time varying phasor depends on the {\em resistive component} of impedance and the quadrature component  depends on the {\em reactive component}. 
The sum of the first two terms in Eq. \eqref{eq:inst3} is active power, and it is determined by the real component of impedance. It may be called the {\em real power} 
that dissipates heat, produces light, and turn shafts. The third term is non-active power, and it is determined by the reactive component of impedance. It may be called the 
{\em reactive power} that accounts for the rate at which energy is exchanged between the electric field that supports voltage and the magnetic field that produces current. 

In summary, active power is real power, and non-active power is reactive power, when current and voltage may be identified in a Thevenin equivalent circuit.

Now in Fig 1b, the 
complex phasor representation of instantaneous power consists of the 
phasor $\frac{ZI^2}{2}e^{j(2\omega_0 t +2\phi)}$ rotating around the tip of the  stationary phasor $\frac{ZI^2}{2}$. The 
rotating phasor
decomposes orthogonally as $(\frac{RI^2}{2}+j\frac{XI^2}{2})e^{j(2\omega_0 t +2\phi)}$. The 
phasor representation of the rotating phasor, namely $(\frac{RI^2}{2}+j\frac{XI^2}{2})$,  has  Cartesian coordinates $( \frac{RI^2}{2} , \frac{XI^2}{2} )$. These are taken 
to define the sides of a right triangle whose hypotenuse  is the {\em apparent power} $\frac{ZI^2}{2}$.  As before, the rotating triangle is 
the {\em power triangle}, the angle $\theta-\phi$ is  the {\em power factor angle}, and $\cos (\theta-\phi)$ is  the {\em power factor}. The power triangle changes orientation, but it does not change shape. 
The instantaneous power is the resolution of the complex result onto the real axis of the Argand plane.  Power is delivered to the impedance load during  $(1-(\theta-\phi)/\pi)T$ seconds of 
each period $T$, and returned to the voltage source during $((\theta-\phi)/\pi)T$ seconds of each period. For these $((\theta-\phi)/\pi)T$ seconds, reactive components are exchanging energy with resistive components at a rate that exceeds the the capacity of the resistive 
components to dissipate heat, produce light or turn shafts. The excess energy is returned to the voltage source, which means the circuit elements are energizing the source, rather than vice-versa. 

\section{The Hilbert Transform, Phase-Splitter, Analytic Signal, and Bedrosian's Theorem}

Suppose we had begun with  
voltage and the current represented as  
 $v(t)=\makebox{Re}\{\tilde{v}(t)\}$, and $i(t)=\makebox{Re}\{\tilde{i}(t)\}$, where $\tilde{v}(t)=Ve^{j\theta}e^{j\omega_0 t}$ and $\tilde{i}(t)=Ie^{j\phi}e^{j\omega_0 t}$ are the {\em complex analytic representations} of real voltage and current. Then we might have defined {\em instantaneous Hermitian power} to be 
\begin{align}
p_{\tilde{v}\tilde{i}^*}(t)=\tilde{v}\tilde{i}^{*}=VIe^{j(\theta-\phi) }
\end{align}
and {\em  complementary instantaneous power} to be 
\begin{align}
p_{\tilde{v}\tilde{i}}(t)=\tilde{v}\tilde{i}=VIe^{j(\theta+\phi)}e^{j(2\omega_0 t+ \theta+\phi)}
\end{align}
Then the complex analytic representation of Eq. \eqref{eq:complexinstpower1} would  have been written as
\begin{align}
p_{vi}(t)=\frac{1}{2}\makebox{Re}\{p_{\tilde{v}\tilde{i}^*}(t)+p_{\tilde{v}\tilde{i}}(t) \}
\end{align}
In Figure 1(b), the fixed phasor may be re-labeled at $(1/2)p_{\tilde{v}\tilde{i}}(t)$ and the rotating phasor as $(1/2)p_{\tilde{v}\tilde{i}^*}(t)$. This suggests that the treatment of instantaneous power might be extended from sinusoidal cases to non-sinusoidal, and in fact aperiodic, signals by appealing to the Hilbert transform and  complex analytic  representations of real signals, as a generalization of the phasor representations of sinusoidal signals.


\subsection{The Hilbert Transform} The Hilbert transform is a linear operator $H$ that transforms real signals $x\in L^2({\cal R})$ into real signals $\hat x\in L^2({\cal R})$ according to the filtering formula

\begin{equation}
\hat x=Hx: \makebox{ } {\hat x}(t)=\int_{-\infty}^{\infty}h(t-u)x(u)du.
\end{equation}

\noindent The real and odd impulse response of the operator $H$ is $h(t), -\infty<t<\infty$, with complex Fourier transform $H(\omega),-\infty<\omega<\infty $:

\begin{equation}
h(t)=\frac{1}{\pi t}\longleftrightarrow \frac{1}{j}sgn(\omega)=H(\omega)
\end{equation}

\noindent The double arrow denotes Fourier transform pair. It is clear that if $x(t)\longleftrightarrow X(\omega)$, then $\hat x(t)\longleftrightarrow sgn(\omega)e^{-j\pi/2}X(\omega)$. So, frequency-by-frequency, the Hilbert transform leaves the magnitude of the signal spectrum unchanged, but changes its phase by $-\pi/2$ for positive frequencies and by $\pi/2$ for negative frequencies. Of course $X(\omega)$, $H(\omega)$, and $\hat X(\omega)$ are Hermitian symmetric functions of radian frequency $\omega$. The filter $H$ is all-pass, but unrealizable. In \cite{Haque02} a simple first-order rational approximation of $H$ is proposed, but in our treatment of generalized power, there is no need for such an approximation. A hardware implementation of $H$ would use a high-order approximation of all-pass $H$. 

The complex signal $\tilde x(t)=x(t)+j\hat x(t)$ is called the analytic version of $x(t)$, or the complex analytic version of $x(t)$, or the complex envelope of $x(t)$. It may be written as the convolution

\begin{equation}
\tilde x=\Psi  x: \tilde x(t)=\int_{-\infty}^{\infty} \psi(t-u)x(u)du
\end{equation}
\noindent where $\psi(t)=\delta(t)+jh(t)$ is the impulse response of the {\em phase splitter} $\Psi$; the Dirac delta $\delta(t)$ is the impulse response of the identity operator  and $h(t)$ is the impulse response of the Hilbert transform. The  Fourier transform of $\psi(t)$ is  easily shown to be $\Psi(\omega)=2step(\omega)$, where $step(\omega)$ is the unit step function. That is, $\psi(t)=\delta(t)+jh(t)\longleftrightarrow 2step(\omega)=\Psi(\omega)$, and $\tilde x(t)\longleftrightarrow 2X(\omega)step(\omega)$. It follows that the spectral representations of real $x,\hat x$ and complex $\tilde x$ are

\begin{align}
x(t)=\int_0^{\infty}2\makebox{Re}\{X(\omega)e^{j\omega t}\}\frac{d\omega}{2\pi}\\
\hat x(t)=\int_0^{\infty}2\makebox{Im}\{X(\omega)e^{j\omega t}\}\frac{d\omega}{2\pi}\\
\tilde x(t)=\int_0^{\infty}2X(\omega)e^{j\omega t}
\frac{d\omega}{2\pi}
\end{align}

These spectrally-efficient representations exploit  the Hermitian symmetry of Fourier transforms for real signals. It is interesting to note that the celebrated Kramers-Kronig relations are duals of these results. That is, the Kramers-Kronig relations establish that the imaginary part of a spectrum must be the Hilbert transform of the real part for the corresponding signal to be causal, whereas these results establish that the imaginary part of a signal must be the Hilbert transform of the real part for its spectrum to be causal.\\

\noindent {\em Example 1}. Consider the real periodic-$\frac{2\pi}{\omega_0}$ signal

\begin{align}
x(t)=&\sum_{-M}^{M}\frac{A_m}{2}e^{j\theta_m}e^{jm\omega_0 t}=\sum_1^{M}A_m\cos(m\omega_0 t+\theta_m),\\
A_0=&0, \, A_{-m}=A_m, \makebox{ and } \theta_{-m}=-\theta_m.
\end{align}
\noindent The Hilbert transform of this signal is $\hat x(t)$, and the complex analytic signal is $\tilde x(t)$:
\begin{align}
\hat x(t)=&\sum_{-M}^{M}\frac{sgn(m)A_m}{2j}e^{j\theta_m}e^{jm\omega_0 t}=\sum_1^{M}A_m\sin(m\omega_0 t+\theta_m),\\
\tilde x(t)=&\sum_1^{M}A_m e^{j(m\omega_0 t+\theta_m)}.
\end{align}
This may be written $\tilde x(t)=A_1e^{j(\omega_0 t+\theta_1)}+\sum_2^{M}A_m e^{j(m\omega_0 t+\theta_m)}$, which is a complex analytic representation of a real sinusoid of frequency $\omega_0$, plus harmonic distortion. A special case of this example is $x(t)=A_1\cos(\omega_0 t+\theta_1)$, in which case the Hilbert transform $\hat x(t)$ and complex analytic signal $\tilde x(t)$ are
\begin{align}
\hat x(t)=&A_1\sin(\omega_0 t+\theta_1)\\
\tilde x(t)=&x(t)+j\hat x(t)=A_1e^{j(\omega_0 t+\theta_1)}=A_1e^{j\theta_1}e^{j\omega_0 t} \end{align}
\noindent Of course, in this case  the complex analytic signal (or complex envelope) ${\tilde x}(t)$ is a {\em rotating phasor} and $A_1e^{j\theta_1}$ is the corresponding {\em stationary phasor}.

Note that, beginning with the complex rotating phasor $Ae^{j\theta}e^{j\omega_0 t}$, the real operator $\makebox{Re}$ returns the real signal $A\cos(\omega_0 t+\theta)$, whereas, beginning with the real signal $A\cos(\omega_0 t+\theta)$, the {\em phase splitter} $\Psi$ returns the complex signal $Ae^{j\theta}e^{j\omega_0 t}$. For this reason we call the phase splitter the {\em unreal operator}. 


\subsection{Bedrosian's Theorem and Its Implications}

Let $x(t)=u(t)v(t)$ be a product of real signals, with $u(t)\longleftrightarrow U(\omega)$ a low-pass signal with $U(\omega)=0$ for $|\omega|>\Omega$, and $v(t)\longleftrightarrow V(\omega)$ a high-pass signal with $V(\omega)=0$ for $|\omega|<\Omega$. Then Bedrosian's theorem shows $\hat x(t)=u(t)\hat v(t)$. That is, the low-pass, slowly-varying, factor may be regarded as constant when calculating the Hilbert transform.\\

\noindent {\em Example 2}. Consider the real aperiodic signal
\begin{equation}
x(t)=A(t)\cos(\omega_0 t+\theta(t))
\end{equation}
\noindent This is a co-sinusoidal carrier of nominal frequency $\omega_0$, amplitude and phase modulated by the lowpass amplitude $A(t)$ and the lowpass phase $\theta(t)$. Then, by Bedrosian's theorem, if the bandwidth of $A(t)\cos(\theta(t))$ is smaller than $\omega_0$, then the Hilbert transform of $x(t)$ is $\hat x(t)$ and its complex analytic signal is $\tilde x(t)$:
\begin{align}
\hat x(t)=&A(t)\sin(\omega_0 t+\theta(t)),\\
\tilde x(t)=&A(t)\cos(\omega_0 t+\theta(t))+jA(t)\sin(\omega_0 t+\theta(t))\\
=&A(t)e^{j\theta(t)}e^{j\omega_0 t}
\end{align}

\noindent In this complex analytic representation of the real signal $x(t)$, the complex envelope $\tilde x(t)$ is a rotating phasor with time-varying amplitude and phase. Among its many virtues, this complex analytic representation of the real signal $x(t)$ provides an unambiguous definition of instantaneous frequency:
\begin{equation}
\omega(t)=\frac{d}{dt}(\omega_0 t+\theta(t))=\omega_0+\frac{d}{dt}\theta(t).
\end{equation}
The complex analytic signal may be demodulated with $e^{-j\omega_0 t}$ to return $A(t)e^{j\theta(t)}$. 
In hardware implementations it is common to approximate the phase splitter-complex demodulator by a {\em quadrature demodulator}, wherein the real signal is multiplied by $\cos(\omega_0 t)$ in the {\em real} channel and by $\sin(\omega_0 t)$ in the quadrature channel; each of these channels is then low-pass filtered.  




\section{Complex Representation of Real-Valued Instantaneous Power for Non-Sinusoidal Signals}
Begin with complex analytic voltage and current, $\tilde{v}(t)$ and $\tilde{i}(t)$, and define the  complex {\em Hermitian  power}, and the complex {\em complementary  power}: 
\begin{align}
p_{\tilde{v}\tilde{i}^*}(t)=&\tilde{v}(t)\tilde{i}^*(t),\\
p_{\tilde{v}\tilde{i}}(t)=&\tilde{v}(t)\tilde{i}(t).
\end{align} 
It is always the case that the instantaneous power is
\begin{align}
p_{vi}(t)=\frac{1}{2}\makebox{Re}\{ p_{\tilde{v}\tilde{i}^*}(t)+
p_{\tilde{v}\tilde{i}}(t)\}
\end{align}

Now consider the  real voltage $v(t)=\makebox{Re}\{V(t)e^{j\theta(t)}e^{j\omega_0 t }\}=V(t)\cos(\omega_0 t+\theta(t))$ and real current $i(t)=\makebox{Re}\{I(t)e^{j\phi(t)}e^{j\omega_0 t} \}=I(t)\cos(\omega_0 t+\phi(t))$. Assume the functions $V(t)e^{j\theta(t)}$ and $I(t)e^{j\phi(t)}$ have bandwidth less than $\omega_0$. Then from Bedrosian's Theorem, it follows that  $\tilde{v}(t)=V(t)e^{j\theta(t)}e^{j\omega_0 t}$
is the complex analytic version of $v(t)$ and $\tilde{i}(t)=I(t)e^{j\phi(t)}e^{j\omega_0 t} $ is the complex analytic version of $i(t)$. These complex analytic signals may be extracted from their real counterparts with the phase-splitter. Then the complex {\em Hermitian  power}, and the complex {\em complementary  power}  compute  as follows:
\begin{align}
\label{eq:slow1}
p_{\tilde{v}\tilde{i}^*}=&(\tilde{v}\tilde{i}^*)(t)=V(t)I(t)e^{j(\theta(t)-\phi(t))}\\
\label{eq:slow2}
p_{\tilde{v}\tilde{i}}=&(\tilde{v}\tilde{i})(t)=V(t)I(t)e^{j(\theta(t)-\phi(t))}e^{j(2\omega_0 t+2\phi(t))}
\end{align} 
The real-valued instantaneous power is
\begin{align}
\label{eq:geninstpower1}
&p_{vi}(t)
=\makebox{Re}\{\frac{V(t)I(t)}{2}e^{j(\theta(t)-\phi(t))}+\frac{V(t)I(t)}{2}e^{j(\theta(t)-\phi(t))}e^{j(2\omega_0 t+2\phi(t))} \}\\
\label{eq:geninstpower2}
=&\makebox{Re}\left\{\frac{V(t)I(t)}{2}e^{j(\theta(t)-\phi(t))}\right.\nonumber\\
+&\left.\frac{V(t)I(t)}{2}[\cos(\theta(t)-\phi(t))+j\sin(\theta(t)-\phi(t))]e^{j(2\omega_0 t+2\phi(t))}\right\}\\
=\label{eq:geninstpower3}
&\frac{V(t)I(t)}{2}\cos(\theta(t)-\phi(t))
+\frac{V(t)I(t)}{2}\cos(\theta(t)-\phi(t))\cos(2\omega_0 t+2\phi(t))\nonumber\\
&~~~~~~-\frac{V(t)I(t)}{2}\sin(\theta(t)-\phi(t))\sin(2\omega_0 t+2\phi(t)) 
\end{align}
Equations \eqref{eq:geninstpower1}-\eqref{eq:geninstpower3} generalize the sinusoidal results of Eqs. \eqref{eq:complexinstpower1}-\eqref{eq:complexinstpower3} to non-sinusoidal signals. These non-sinusoidal signals need not be periodic.The first term in Eqs. \eqref{eq:geninstpower1} and \eqref{eq:geninstpower2} may be called slowly-varying, or short term average power. 
The second and third terms in Eq. \eqref{eq:geninstpower2}  resolve the second term in Eq. \eqref{eq:geninstpower1} into its in-phase and quadrature components. Eq. \eqref{eq:geninstpower3} resolves the real operator to show that instantaneous power is slowly-varying average power plus the in-phase and quadrature components of the component of instantaneous power that oscillates at high frequency $2\omega_0$. Moreover, the sum of the first two terms in Eq. \eqref{eq:geninstpower3} accounts for active power and the third term accounts for non-active power. In a time-varying linear system, active power is real power and non-active power is reactive power.

The geometry generalizes the geometry of Fig. 1(b): The complex representation of instantaneous power consists of the slowly varying phasor $\frac{V(t)I(t)}{2}e^{j(\theta(t)-\phi(t)}$, around which rotates the phasor $\frac{V(t)I(t)}{2}e^{j(\theta(t)-\phi(t))}$. 
This rotating phasor decomposes into its in-phase and quadrature components $\frac{V(t)I(t)}{2}\cos(\theta(t)-\phi(t)))$ and $ \frac{V(t)I(t)}{2}\sin(\theta(t)-\phi(t))$. This decomposition  shows that a slowly-flexing right triangle, with sides equal to these in-phase and quadrature components, rotates rapidly around the tip of a slowly-varying phasor. This triangle may be called a slowly-varying power triangle.

\section{Interpretations in a Time-Varying Thevenin Equivalent Circuit}

When can active power and non-active power be called real power and reactive power in the case where voltage and current are non-sinusoidal and the impedance is time-varying? 

Begin with a Thevenin equivalent circuit for which the  impedence and corresponding impulse response are time-varying. That is, the impedance $Z(t,\omega)$ is the Fourier transform of the real time-varying impulse response $z(t,\tau): Z(t,\omega)=\int z(t,\tau)e^{-j\omega \tau}d\tau$. Assume the real current is $i(t)=I\cos(\omega_0 t+\phi)$, with corresponding complex analytic representation $\tilde{i}(t)=Ie^{j(\omega_0 t+\phi)}$. The real voltage is  
\begin{align}
v(t)=&\int \,z(t,\tau) i(t-\tau)d\tau=\makebox{Re}\{Z(t,\omega_0)Ie^{j\phi}e^{j\omega_0 t} \}.
\end{align}
where $z(t,\tau)\longleftrightarrow Z(t,\omega)$. 
Assuming that the bandwidth of $Z(t, \omega_0)$ is less than $\omega_0$, Bedrosians theorem shows that the complex  analytic voltage is  
\begin{align}
\tilde{v}(t)=Z(t,\omega_0 )Ie^{j(\omega_0 t+\phi)}
\end{align}
With these results, the Hermitian, complementary, and real-valued instantaneous power are
\begin{align}
\label{eq:slowimped1}
&p_{\tilde{v}\tilde{i}^*}(t)=\tilde{v}(t)\tilde{i}^*(t)=Z(t,\omega_0)I^2=R(t)I^2+jX(t,\omega_0)I^2,\\
\label{eq:slowimped2}
&p_{\tilde{v}\tilde{i}}(t)=\tilde{v}(t)\tilde{i}(t)
=Z(t,\omega_0)I^2e^{j(2\omega_0 t+2\phi)} \\
\label{eq:slowimped3}
&p_{vi}(t)=\frac{1}{2}\makebox{Re}\{p_{\tilde{v}\tilde{i}}(t)+ p_{\tilde{v}\tilde{i}^*}(t)\}\nonumber \\
&=\frac{R(t)I^2}{2}+\frac{R(t)I^2}{2}\cos(2\omega_0 t+2\phi)-\frac{X(t, \omega_0)I^2}{2}\sin(2\omega_0 t+2\phi).
\end{align}
As before, Figure 1(b) may be re-labeled, in this case with time-varying impedance components. The sum of the first two terms in Eq. \eqref{eq:slowimped3} is active power, and it is real power. The third term is non-active power, and it is reactive power. 

Can this analysis be extended  to more general currents? It does not seem possibe. But this example is useful. It assumes a regulated current source, and a time-varying impedance. And of course by letting $v(t)$ be sinusoidal, and allowing admittance to be time varying, this example applies to a regulated voltage source, and a time-varying admittance. 

\begin{figure*}[t!]
  \centering
  \includegraphics[width=\linewidth]{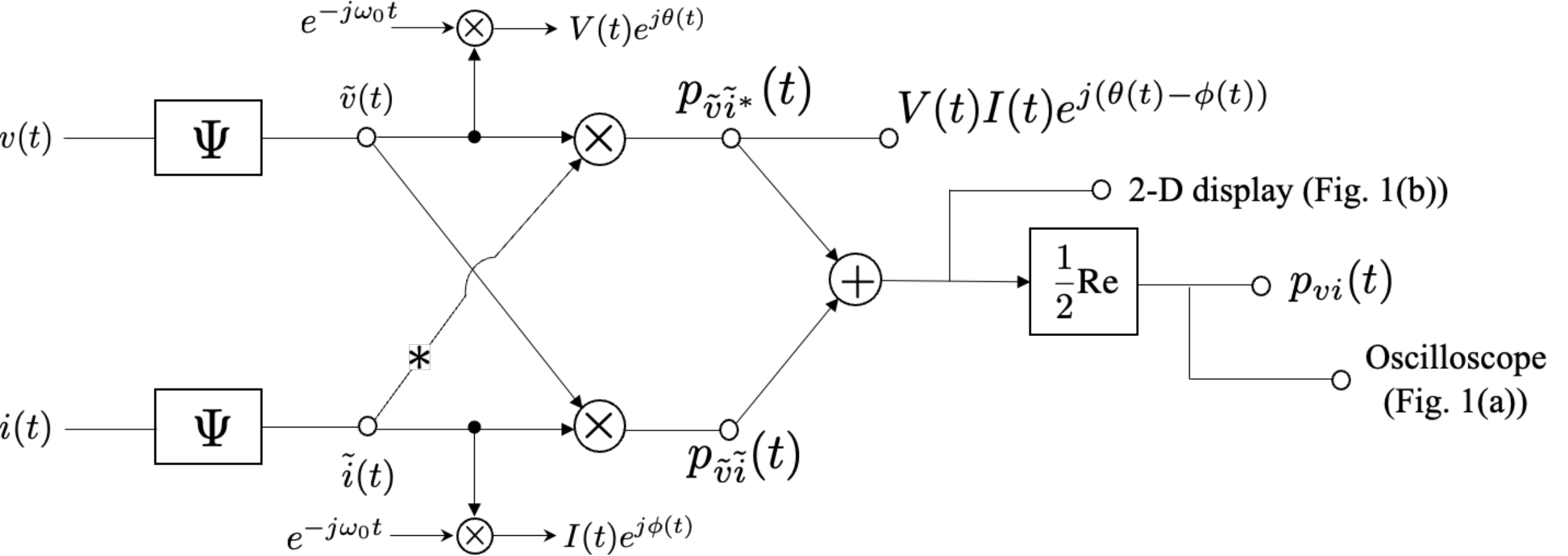}
  \caption{\small Hardware diagram for extracting and displaying components of Hermitian, complementary, and real-valued instantaneous power.}
  \label{fig:meter}
\end{figure*}

\section{Spectral Theory of Average Power}
Define average power to be a time average.  For sinusoidal voltage and current the results are straight forward:
\begin{align}
&P_{\tilde{v}\tilde{i}}=\int p_{\tilde{v}\tilde{i}}(t)dt=\frac{VI}{2}e^{j(\theta-\phi)},\\
&P_{\tilde{v}\tilde{i}^*}=\int p_{\tilde{v}\tilde{i}^*}(t)dt=0,\\
&P_{vi}=\frac{VI}{2}\cos(\theta-\phi).
\end{align}
There is no need for a spectral representation of these results.
For general time-varying voltage and current, spectral representations bring insight, if correctly interpreted. 

Begin with the Fourier transform pairs $v(t)\longleftrightarrow V(\omega)$,
$i(t)\longleftrightarrow I(\omega)$. The current and voltage are real, but their Fourier transforms are complex, with Hermitian symmetry. Give the complex Fourier transforms the polar representations $V(\omega)=A(\omega)e^{j\Theta(\omega)}$ and $I(\omega)=B(\omega)e^{j\Phi(\omega)}$. Recall the corresponding one-sided spectral representations for analytic signals have a factor of $2$. Then it is straightforward to  write time domain averages as the following frequency-domain averages:
\begin{align}
P_{\tilde{v}\tilde{i}^*}=&\int _0^{\infty}4V(\omega)I^*(\omega)\frac{d\omega}{2\pi}=\int_0^{\infty} 4A(\omega)B(\omega)e^{j(\Theta(\omega)-\Phi(\omega))}\frac{d\omega}{2\pi}\\
P_{\tilde{v}\tilde{i}}=&\int_0^{\infty} 4V(\omega)I(\omega)\frac{d\omega}{2\pi}=\int_0^{\infty} 4A(\omega)B(\omega)e^{j(\Theta(\omega)+\Phi(\omega))}\frac{d\omega}{2\pi}\\
=&\int_0^{\infty} 4A(\omega)B(\omega)\big [\cos(\Theta(\omega)-\Phi(\omega))\nonumber \\
+&j\sin(\Theta(\omega)-\Phi(\omega))\big]e^{j2\Phi(\omega)}\frac{d\omega}{2\pi}\\
P_{vi}
\label{eq:freqinst3}
=&\int_0^{\infty} \big[2 A(\omega)B(\omega)\cos(\Theta(\omega)-\Phi(\omega))\nonumber \\
+& 2A(\omega)B(\omega)\cos(\Theta(\omega)-\Phi(\omega))\cos(2\Phi(\omega)) \nonumber\\
-& 2A(\omega)B(\omega)\sin(\Theta(\omega)-\Phi(\omega))\sin(2\Phi(\omega))\big]\frac{d\omega}{2\pi}
\end{align}
The sum of the first two terms is active power and the third term is non-active power.

Frequency-by-frequency, these average powers have the same components as the 
formulas in the sinusoidal case. 
The geometry of the complex representation of average power is the geometry of the Hermitian phasor $A(\omega)B(\omega)e^{j(\Theta(\omega)-\Phi(\omega))}$ plus the complementary phasor $A(\omega)B(\omega)e^{j(\Theta(\omega)-\Phi(\omega))}e^{j2\Phi(\omega)}$;  the complementary   phasor resolves as $4A(\omega)B(\omega)\big [\cos(\Theta(\omega)-\Phi(\omega))
+j\sin(\Theta(\omega)-\Phi(\omega))\big]e^{j2\Phi(\omega)}$. So frequency-by-frequency, it remains the case that  $4A(\omega)B(\omega)\cos(\Theta(\omega)-\Phi(\omega))
+j4A(\omega)B(\omega)\sin(\Theta(\omega)-\Phi(\omega)$ is  rotated by $e^{j2\Phi(\omega)}$. 
Frequency-by-frequency, one may speak of a  power triangle whose hypotenuse has magnitude equal to the frequency-dependent apparent power, $4A(\omega)B(\omega)$, and whose sides are $2A(\omega)B(\omega)\cos(\Theta(\omega)-\Phi(\omega))$ and $2A(\omega)B(\omega)\sin(\Theta(\omega)-\Phi(\omega))$. {\em It cannot be said that there is a broadband rotating triangle}. That is, because the square of a sum is not a sum of squarees,
\begin{align}
&\left(\int 2A(\omega)B(\omega) \frac{d \omega}{2\pi}\right)^2  \nonumber\\&\neq\int\left(2A(\omega)B(\omega)\right)^2 \left(\cos^2(\Theta(\omega)-\Phi(\omega)\right)+\sin^2\left(\Theta(\omega)-\Phi(\omega)\right)\frac{d \omega}{2\pi} \\
&=\int \left(2A(\omega)B(\omega)\right)^2 \cos^2\left(\Theta(\omega)-\Phi(\omega)\right)\frac{d \omega}{2\pi}+\nonumber\\ &~~~~~~~~~~~~~\int \left(2A(\omega)B(\omega)\right)^2 \sin^2\left(\Theta(\omega)-\Phi(\omega)\right)\frac{d \omega}{2\pi}
\end{align}
This is the basis of Czarnecki's objection \cite{Cza1985} to Budeneu's definition of reactive power in non-sinusoidal situations \cite{Bud1927}. 

If $V(\omega)$ and $I(\omega)$ are voltage and current in a Thevenin equivalent circuit, then it is straight forward to write $V(\omega)=Z(\omega)I(\omega)$, in which case $V(\omega)I^*(\omega)=Z(\omega)|I(\omega)|^2$ and  $V(\omega)I(\omega)=Z(\omega)I^2(\omega)$. If the impedance $Z=R(\omega)+jX(\omega)$ is written as $Z(\omega)=|Z(\omega)|e^{j(\Theta(\omega)-\Phi(\omega))}$, then $R(\omega)=|Z(\omega)|\cos(\Theta(\omega)-\Phi(\omega))$ and $X(\omega)=|Z(\omega)|\sin(\Theta(\omega)-\Phi(\omega))$. Then the frequency domain  formulas may be written
\begin{align}
&P_{\tilde{v}\tilde{i}^*}(\omega)=4|Z(\omega)||I(\omega)|^2e^{j(\Theta(\omega)-\Phi(\omega))}\\
&P_{\tilde{v}\tilde{i}}(\omega)=4|Z(\omega)||I^2(\omega)|e^{j(\Theta(\omega)-\Phi(\omega))}e^{j2\Phi(\omega)}\nonumber \\
&=4|Z(\omega)|||I(\omega)|^2\times\nonumber\\ &~~~~~~\big[(\cos(\Theta(\omega)-\Phi(\omega))\!+\!j\sin(\Theta(\omega)-\Phi(\omega))\big]e^{j2\Phi(\omega)}\\
&P_{vi}(\omega)=2R(\omega)||I(\omega)|^2+2R(\omega)|I|^2(\omega)\cos(2\Phi(\omega))\nonumber \\
&~~~~~~~~~~~~~~~~~-2X(\omega)||I(\omega)|^2\sin(2\Phi(\omega)) 
\end{align}
So, \textit{frequency-by-frequency}, active power depends on real impedance, and non-active power depends on reactive impedance. Active power is real power and non-active power is reactive power.  Of course the formula for $P_{vi}(\omega)$ may be integrated to get average power as average real power plus average reactive power.

This frequency domain analysis goes through essentially unchanged for discrete spectra $\{V_n, n=0,\pm 1, \ldots \}$, and $\{I_n, n=0,\pm 1, \ldots \}$, in which case these line spectra would be describing non-sinusoidal periodic signals. 


\section{Hardware Diagrams for Extracting Components of Instantaneous Power}

The results so far establish the components of Hermitian, complementary, and real-valued instantaneous power. They do not show how these components are to be extracted for monitoring of power system performance. 

Begin with the real voltage $v(t)$ and current $i(t)$, measured by volt-meter and current transformer. As illustrated in the hardware diagram of Fig. \ref{fig:meter}, these are Hilbert transformed to produce the analytic voltage $\tilde{v}(t)$ and analytic current  $\tilde{i}(t)$. From these are computed the Hermitian and complementary powers $p_{\tilde{v}\tilde{i}}$ and $p_{\tilde{v}\tilde{i}^*}$. The sum of these may be displayed for the complex representation of real-valued instantaneous power. Or, the complementary power may be demodulated to baseband as $e^{-j2\omega_0 t}p_{\tilde{v}\tilde{i}^*}$. This stationary picture is the picture of Fig. 1b at $t=0$. If the voltage and current are slowly-varying, the picture will be the picture of Eqs. \eqref{eq:slow1} and \eqref{eq:slow2}. If the voltage and current are constrained by the impedance of a Thevenin equivalent circuit, the the picture will be the picture of Eqs. \eqref{eq:slowimped1}-\eqref{eq:slowimped3}. For computed quantities, the real and imaginary parts of the Hermitian power return $V(t)I(t)\cos(\theta(t)-\phi(t)$ and $V(t)I(t)\sin(\theta(t)-\phi(t)$. These determine the three terms of real-valued power, provided $\omega_0$  and $\phi$ are known; $\omega_0$ is assumed known, or may be extracted with a frequency estimator, and $\phi$ is determined from the identity $e^{-j\omega_0 t}p_{\tilde{v}\tilde{i}}=e^{j2\phi}p_{\tilde{v}\tilde{i}}$. 

\section{Conclusions}
Hermitian and complementary complex power may be computed from the Hilbert transforms of real  voltage and current. These, in turn, provide an evocative complex representation  of real-valued instantaneous power. This representation extends to non-sinusoidal cases, and when there is a Thevenin equivalent circuit, the complex representation may be used to identify active and non-active powers as real and reactive powers. 
A simple hardware diagram may be used to extract these components. Frequency domain formulas for average power decompose in a way that is dual, frequency-by-frequency, to the time domain formulas for instantaneous power.

\bibliographystyle{IEEEtran}

\end{document}